# Universality of Polymer Dynamics near the Glass Transition and the Relationship between the Simulated and Experimental Glass Transition Temperatures of Amorphous Polymers

Valeriy V. Ginzburg[1,*], Oleg V. Gendelman[2], Riccardo Casalini[3], and Alessio Zaccone[4]

[1]Department of Chemical Engineering and Materials Science, Michigan State University, East Lansing, Michigan, USA 48824

[2]Faculty of Mechanical Engineering, Technion, Haifa 32000003, Israel

[3]Chemistry Division, Naval Research Laboratory, 4555 Overlook Avenue SW, Washington, D.C., USA 20375

[4]University of Milan, Department of Physics, via Celoria 16, 20133 Milano, Italy

*Corresponding author, email ginzbur7@msu.edu

# TOC Graphic

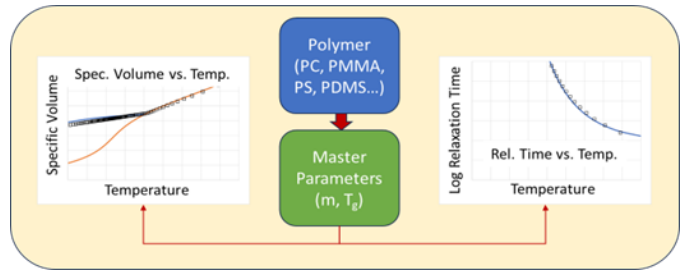

# Abstract

Describing the dynamics and thermodynamics of amorphous materials near the glass transition is a major challenge in soft-matter physics and polymer engineering. Here, we show that the dependence of the dielectric $\alpha$-relaxation time on temperature can be captured by a universal equation with only two independent parameters, $T_g$ and fragility, $m$. This is similar in spirit to the ideas of van Krevelen and Bicerano, and can be related to the Boyer-Spencer and Simha-Boyer rules for the volumetric thermal expansion via a modified free-volume approach such as Sanchez-Lacombe "Two-state, two-(time)scale" (SL-TS2) theory. The model is compared to experimental data for nine amorphous polymers with varying values of $T_g$ and $m$, and a good qualitative and quantitative agreement is found. We also derive the relationship between the experimental and computational (high-cooling-rate) $T_g$, and compare our model prediction for the $T_g$-shift between the two with the simulation results of Afzal et al., finding a good qualitative and semi-quantitative agreement. The results could serve as a guidance for future simulation studies.

# 1. Introduction

Glass transition in polymers is a complex and still poorly understood phenomenon that impacts many important material properties, from viscosity and elastic modulus to thermal and electrical conductivity to gas permeability.[1–3] It is thus desirable to have simple and reliable thermodynamic and dynamic models describing the behavior of thermodynamic (density, heat capacity) and dynamic (viscosity, relaxation time) properties as functions of external conditions (temperature, pressure) and processing history (heating or cooling rate, etc.) While significant progress has been achieved over the years, many challenges remain.

While the question of choosing the right model for the description of the glass transition is important, an even more interesting question is related to the "universality" or "the law of corresponding states". Unlike crystalline solids, liquids (including polymers) are difficult to describe with a simple equation of state (for a recent discussion on the topic, see Proctor and Trachenko[4]). Specifically, we can formulate our question as follows. Suppose we want to describe the specific volume (the inverse of the density) and the $\alpha$-relaxation time as functions of temperature, $T$, and pressure, $P$. We could define a timescale, $\tau_0$, a volume scale, $v_0$, and an energy scale, $\varepsilon_0$. Then, a combined dynamic-thermodynamic model would have these three scale-setting parameters and, in addition, one or more material-dependent dimensionless parameters (fragility being one example). We need to understand how many of these parameters are truly independent and how many are interrelated. To do this, we specifically divide the problem into two parts, as discussed below.

The first problem is the isobaric volume expansion in the glassy and liquid state. It has been observed by Boyer *et al.* [5,6] that the coefficients of thermal expansion (CTE) for many polymers are very close when multiplied by the glass transition temperature. In particular, $\alpha_L T_g \approx 0.19 \pm 0.03$ (Boyer-Spencer rule)[5] and $(\alpha_L - \alpha_G)T_g \approx 0.115$ (Simha-Boyer rule),[6] where $\alpha_L$ and $\alpha_G$ are the isobaric volumetric coefficients of thermal expansion in the liquid and glassy states, respectively. The Boyer-Spencer rule also emerged (along with the Flory-Fox dependence on the molecular weight) within a molecular-level theory proposed by Zaccone and Terentjev.[7] More recently, Lunkenheimer et al.[8] suggested a different scaling, $\alpha_L T_g/m$ = const., where $m$ is the dynamic fragility (to be defined below). The new scaling was suggested to be universal for both polymeric and non-polymeric glass-formers; if one restricts the analysis to polymers only, the two scaling rules work equally well. The Boyer-Spencer rule would imply that if we introduce new dimensionless "corresponding state" variables, $x = T/T_g$ and $y = (V[T]/V_g)$, there should be no additional material-dependent parameters (apart from the specific volume at the glass transition, $V_g$).

Second, let us consider the isobaric (atmospheric pressure) dependence of the relaxation time (or viscosity) on temperature. Various experiments – broadband dielectric spectroscopy, viscometry, dynamical mechanical spectroscopy, nuclear magnetic resonance, and others – are consistent with the "super-Arrhenius" behavior phenomenologically described by the Vogel-Fulcher-Tammann-Hesse (VFTH)[9–11] or Williams-Landel-Ferry (WLF),[12] $log(\tau) = log(\tau_\infty) + \frac{A}{T-T_0}$ (VFTH) or $log(\tau) = log(\tau_{ref}) - \frac{C_1(T-T_{ref})}{C_2+(T-T_{ref})}$ (WLF). (Here and below, "log" means base-10 logarithm, and relaxation time is measured in seconds). These equations are equivalent, and can be derived based on free volume theory (FVT)[13] or the configurational entropy (Adam-Gibbs) theory.[14] Within this framework, the relaxation time diverges as temperature approaches the so-called "Kauzmann temperature"[15] (labeled $T_0$ in the VFTH expression or $T_{ref} - C_2$ in the WLF expression); the Kauzmann temperature is usually about 50-60 K below the glass transition temperature. Other, more complex, equations have also been proposed (MYEGA,[16] Avramov-Milchev,[17] Elmatad et al. [ECG],[18] Ginzburg [TS2],[19] Laukkanen and Winter,[20] Krausser-Samwer-Zaccone [KSZ],[21] and others). The need to use more complex functional forms stems from observations by McKenna and co-workers that the apparent activation energy probably does not diverge at any finite temperature[22–24] and the (possibly related) comments by Sokolov and Novikov[25] about the non-monotonicity of the second derivative of $\log(\tau)$ with respect to temperature. Sokolov and Novikov suggested that a correct parameterization of the relaxation time must have more than three parameters, and pointed out the Cohen-Grest[26] equation as one example.

Regardless of the number of parameters in the model, we can always re-formulate it in terms of "corresponding states", i.e., *x = T/T$_g$* and *y = log(τ[T]/τ$_g$)*, where as usual, $\log(\tau_g) = 2.0$, and $T_g$ is the temperature for which *τ[T$_g$] = τ$_g$ = 100 s*. In this case, if the model has *M* parameters (*M* = 3 for WLF, AM, MYEGA, *4* for CG, 5 for TS2, etc.), then two parameters are setting the temperature and time scales, and the remaining *(M-2)* can be expressed as dimensionless numbers. The question is – how many of these *(M-2)* parameters are independent and how many are "universal"? Here, we hypothesize that for many polymers, all these parameters must be universal, at least in the region where WLF equation holds. This is demonstrated by showing that the curves for different polymers can be superimposed onto a single master curve by appropriate vertical and horizontal shifts. Notably, such a universality for the dependence of polymer viscosity on the reduced temperature has been proposed, e.g., by Wang and Porter,[27] Ding and Sokolov,[28] Bailly et al.,[29] van Krevelen[2] and Bicerano.[3]

Once the scaling and universality is demonstrated, it is necessary to develop the right theory. Many existing frameworks, including LCL-CFV of Lipson and White,[30–33] the generalized entropy theory of Douglas, Freed, and co-workers,[34–36] the ECNLE theory of Schweizer *et al.*[37–40] and others, can describe both the relaxation time and the density dependence on the temperature. Thus, in our view, any of those theories can be adapted to describe both the universality of the relaxation time profiles and the Boyer rules for thermal expansion. Here, we demonstrate such a universal parameterization using the "Sanchez-Lacombe Two-State, Two (time) Scale" (SL-TS2) theory.[41,42] We regress the five SL-TS2 parameters, use them to compute the master curve, and approximate it with a VFTH equation. This allows one to compute, for example, the WLF coefficients $C_1$ and $C_2$ as functions of $T_g$ and dynamic fragility, $m$.

The proposed scaling and universality rules also help one to model a very important question – how the experimental $T_g$ is related to one obtained in atomistic Molecular Dynamics (MD) simulations. Since the simulation timescale is about 10-12 orders of magnitude smaller than experimental one, it is necessary to come up with theoretical or empirical relationship between the two, as discussed by multiple authors.[43–52] Here, we show that the accuracy and precision of this relationship depends on the polymer fragility and that for low-fragility ("strong") polymers, such prediction is likely to suffer from poor accuracy and precision.

The paper is structured as follows. In Section 2, we describe the dielectric relaxation data and the polymers chosen for the study, as well as the proposed scaling approach and the outlines of the SL-TS2 model. In Section 3, we compare model fits with experiments for the nine polymers chosen for this study, and investigate the relationship between the simulated and experimental $T_g$, concluding with a short discussion. Section 4 contains summary and conclusions.

## 2. Materials, Methods, and Models

### *2.1. Dielectric Relaxation Data*

Broadband dielectric spectroscopy (BDS) is typically utilized to characterize the polymer dynamics. Generally, the lowest-frequency loss-modulus ($\varepsilon''$) peak in the dielectric spectrum is labeled $\alpha$-peak, and the inverse of the frequency corresponding to this peak is labeled $\alpha$-relaxation time or $\tau_\alpha$. Typically, the BDS frequency range is ~$10^7 - 10^{-2}$ Hz,[53] which translates to the time range ~ $10^{-8} - 10^1$ s (i.e., the upper limit is close to the glass transition point which is $\tau_g = 10^2 s$). The dependence of $\tau_\alpha$ on the temperature, $T$, in the vicinity of the glass transition point, is usually well-parameterized by the VFTH[9–11] or WLF[12] equation. It is often convenient to re-parameterize those equations in the form proposed by Rössler and co-workers ("Blochowicz equation"),[54,55]

$$\log\left(\frac{\tau_\alpha(T)}{\tau_\infty}\right) = \frac{K_0^2}{m\left(\frac{T}{T_g}-1\right)+K_0} \qquad (1)$$

Here, $K_0 = 2 - \log(\tau_\infty)$, and $m$ is the dynamic fragility,

$$m = \frac{d\,\log(\tau_\alpha)}{d\left(\frac{T_g}{T}\right)}\bigg|_{T=T_g} \qquad (2)$$

In Table 1, the Blochowicz equation parameters are summarized for several amorphous polymers.

*Table 1. Dynamic parameters for selected amorphous polymers*

| Polymer Name | $T_g$, K | m | log($\tau_\infty$, s) |
|---|---|---|---|
| PDMS | 144 | 126 | -15.5 |
| PBD | 180 | 75 | -11.5 |
| PIB | 198 | 35 | -12.5 |
| PVAc | 307 | 75 | -13.1 |
| PVC | 350 | 166 | -12.1 |
| PCHMA | 354 | 62 | -12.3 |
| PS | 373 | 101 | -11.6 |
| PMMA | 378 | 100 | -10.3 |
| PC | 423 | 99 | -10.2 |

The parameters are compiled from the following literature sources: PDMS,[56] PBD,[56] PIB,[57] PVAc,[58] PVC,[59] PCHMA,[60] PS,[56] PMMA,[61] and PC.[62] Other researchers published slightly different parameterizations of the dielectric data for the same polymers – as discussed above, there could be a significant variability based on the polymer molecular weight and molecular weight distribution. We will address this topic in a future paper.

## 2.2. The Horizontal and Vertical Scaling for the Relaxation Time

To demonstrate the universality of the relaxation time behavior for polymers near $T_g$, the following procedure is utilized. This analysis is similar to the one discussed by Bailly et al.,[29] but with some differences that will be discussed later. First, the separate data for various polymers are plotted in a standard way as $log(\tau_\alpha)$ vs. $T/T_g$. Next, the plot is transformed from linear-log to log-log, i.e., $log(\tau_\alpha)$ vs. $log(T/T_g)$. After that, horizontal and vertical shifts are utilized to collapse all curves together. Finally, the X-axis is transformed back to linear scale.

The above transformation allows one to demonstrate, as we will show later (Figure 1), that there is a master curve described as follows,

$$log\left(\frac{\tau_\alpha(T)}{\tau_{el}}\right) = \mathbb{F}\left(\frac{T}{T_x}\right) \tag{3}$$

where $\mathbb{F}(x)$ is some function to be determined later. The new parameters, $\tau_{el}$ and $T_x$, are related to the magnitude of the shifts and are given by,

$$\frac{T_x}{T_g} = g(m) \tag{4a}$$

$$log(\tau_{el}) = 2 - \mathbb{F}\left(\frac{1}{g(m)}\right) \tag{4b}$$

The functions $\mathbb{F}$ and $g$ are given by,

$$\mathbb{F}(x) = \frac{k}{x-b} \tag{5a}$$

$$g(m) = \frac{[k+mb] - \sqrt{[k+mb]^2 - 4m^2b^2}}{2b^2m} \tag{5b}$$

Equation 5a is simply an approximate functional form to allow an easy VFTH or WLF parameterization. Equation 5b follows from equation 5a and the definition of $m$ (equation 2).

The master curve (equation 3) can be explained based on the free-volume theory and the idea of corresponding states.[2,3,27–29] This is discussed next.

*2.2. The "Universal" Sanchez-Lacombe Two-state, two-(time)scale (SL-TS2) Model*

The Sanchez-Lacombe two-state, two (time)scale (SL-TS2) theory has been described in several earlier publications,[19,41,42,63,64] so here we only recap the major points. The SL-TS2 theory is essentially a combination of a two-state lattice model and a free-volume theory. It is stipulated that a glass-forming, amorphous material can exist in two forms, the high-entropy, high-energy liquid ("L" or "1") and the low-entropy, low-energy glassy solid ("S" or "2"). The material is split into cooperatively rearranging regions (CRR). It is further assumed that within each CRR, a small fraction of the total volume is taken by the "free volume" or "voids". The fraction of CRRs in the "solid" state is labeled $\psi$ and the volume fraction of non-void lattice elements (the "occupancy") is labeled $v$; thus, the fractional free volume (FFV) is simply (1-$v$). Adopting the lattice model framework of Sanchez and Lacombe (SL)[65–68] and labeling the volume of one lattice element as $v_0$, we can write the total CRR volume as,

$$V_{CRR} = v_0[v[\psi r_S + (1-\psi)r_L] + (1-v)] = v_0 \frac{r}{v} \qquad (6)$$

where $r = \psi r_S + (1-\psi)r_L$, $r_S$ and $r_L$ are the number of lattice sites occupied by an all-solid or all-liquid CRR in the absence of voids. The free energy per CRR is written as,

$$G = k_B T \left[ \frac{V_{CRR}}{v_0} \right] \left[ \frac{\phi_S}{r_S} \ln \phi_S + \frac{\phi_L}{r_L} \ln \phi_L + \phi_V \ln \phi_V \right]$$

$$- \frac{z}{2} \left[ \frac{V_{CRR}}{v_0} \right] \left[ \varepsilon_{SS} \phi_S^2 + 2\varepsilon_{SL} \phi_S \phi_L + \varepsilon_{LL} \phi_L^2 \right] \qquad (7)$$

Here, $z$ is the lattice coordination number, $\varepsilon_{SS}$, $\varepsilon_{SL}$, and $\varepsilon_{LL}$ are the interaction ("bonding") energies between adjacent lattice sites of different kind (S-S, S-L, and L-L "bonds"). Other model variables and the details of the free energy minimization procedure are described in Supporting Information. For the equilibrium state, we can minimize the free energy (equation 7) with respect to variables $\psi$ and $\nu$. The equilibrium solutions (see below for more details) are labeled $\psi_{eq}[\tilde{T}]$ and $\nu_{eq}[\tilde{T}]$. Here the dimensionless temperature $\tilde{T} = \frac{k_B T}{\varepsilon^*}$, with $\varepsilon^* = \frac{z\varepsilon_{SS}}{2}$. We can also define dimensionless interaction energies, $\alpha_{LL} = \frac{\varepsilon_{LL}}{\varepsilon_{SS}}$, and $\alpha_{SL} = \frac{\varepsilon_{SL}}{\varepsilon_{SS}}$, and assume that they are related via the Berthelot "geometric mean" rule, $\alpha_{SL} = \sqrt{\alpha_{LL}}$. Finally, we will define the "average" CRR size, $\bar{r} = 0.5(r_S + r_L)$, and the relative volume difference, $\xi = 0.5(r_L - r_S)/\bar{r}$.

The Johari-Goldstein[69,70] $\beta$-relaxation time and the segmental $\alpha$-relaxation time are given by,

$$\tau_\beta[\tilde{T}] = \tau_{el}\, exp\left[\frac{\widetilde{E_1}}{\tilde{T}}\right] \quad (7a)$$

$$\tau_\alpha[\tilde{T}] = \tau_{el}\, exp\left[\frac{\widetilde{E_1}}{\tilde{T}} + \frac{\widetilde{E_2}-\widetilde{E_1}}{\tilde{T}}\psi_{eq}[\tilde{T}]\right] \quad (7b)$$

where $\widetilde{E_1}$ and $\widetilde{E_2}$ are dimensionless activation energies (see Supporting Information), and $\tau_{el}$ is the "elementary" time. Note that in earlier papers, we used the notation $\tau_\infty$ -- here, $\tau_{el}$ is used to distinguish from the parameter $\tau_\infty$ in the VFTH parameterization of the relaxation time.

The non-equilibrium dynamics are described based on the assumption that the solid fraction, $\psi$, relaxes over the timescale $\tau_\beta$, and the occupancy, $\nu$, over the timescale $\tau_\alpha$,

$$\frac{d\psi}{dt} = \frac{\Psi^*[\nu,\tilde{T}]-\psi}{\tau_\beta[\tilde{T}]} \quad (8a)$$

$$\frac{d\nu}{dt} = \frac{\nu_{eq}[\tilde{T}]-\nu}{\tau_\alpha[\psi,\tilde{T}]} \quad (8b)$$

Here, $\Psi^*[v,\tilde{T}]$ is the solution of equation $\frac{\partial G[\psi,v;\tilde{T}]}{\partial \psi} = 0$. Obviously, eqs 8a-b describe isothermal relaxation; to capture the cooling process, one needs to incorporate the temperature evolution as well. The detailed numerical procedures used to compute the equilibrium and constant-cooling-rate solutions of the SL-TS2 model are provided in Supporting Information.

Equation 7 describes the dependence of the material volume on the temperature (via the temperature-dependent variables ψ and v). Such a dependence is known as "equation of state" (EoS); typically, an EoS describes the dependence of the specific volume on temperature T and pressure P, but here we only consider a cross-section in the plane P = 0.1 MPa. (Note that the pressure dependence within SL-TS2 is captured via a separate scaling – see Ginzburg et al.[41,42] and Supporting Information). Over the years, many equations of state have been formulated for polymers.[65–67,71–81] In many cases, equations of state can be written in "universal" form, i.e., *(V/V\*) = f((T/T\*),(P/P\*))*, where *f* is a universal function and *V\*, T\*,* and *P\** are characteristic specific volume, temperature, and pressure of an individual polymer ("the law of corresponding states").

To understand how the law of corresponding states could be incorporated within SL-TS2 framework, we recall the approximate scaling relationships proposed by Simha and Boyer[6] and Boyer and Spencer[82]

$$\alpha_L T_g \approx 0.19 \pm 0.03 \tag{9a}$$

$$[\alpha_L - \alpha_G]T_g \approx 0.115 \tag{9b}$$

Here, $\alpha_L$ and $\alpha_G$ are volumetric coefficients of thermal expansion of a polymer in liquid and glassy state, respectively. For more discussion on these and other relationships, see, e.g., van Krevelen.[2] More recently, Samwer, Zaccone, and co-workers[8,21] analyzed the data for thermal expansion in many polymeric and non-polymeric materials and suggested a rule ($\alpha_L/\alpha_G$) ≈ 2.5—3.5, which is roughly consistent with the Simha-Boyer and Boyer-Spencer rules. We note that in general, quantifying the accuracy and precision of equations 9a-9b is complicated because: (i) the coefficients of thermal expansion are themselves (weakly) temperature-dependent; (ii) the signal-to-noise ratio is relatively high for the glassy state of the material, and (iii) the glassy CTE depends, at least somewhat, on the cooling rate and sample history. Despite these limitations, we will consider equations 9a-9b as our "data points" when constructing and parameterizing a "universal" equation of state for polymers.

One difficulty with equation 9a is as follows. The coefficient $\alpha_L$ is an equilibrium thermodynamic property, while $T_g$ is a function of cooling or heating rate. We thus propose to modify the above relationships as follows,

$$\alpha_L T_x \approx 0.19 \pm 0.03 \tag{10a}$$

$$[\alpha_L - \alpha_G]T_x \approx 0.115 \tag{10b}$$

The temperature $T_x$ introduced here is the same as in Section 2.2. We now stipulate its physical meaning in the context of the SL-TS2 theory – it is the temperature at which the numbers of liquid and solid CRRs in the material are equal, i.e., $\psi$ = 0.5. This means that $T_x$ is an infinite-order (non-singular) mean-field phase transition between solid and liquid, of which the glass transition is a dynamic manifestation. With the above modification, we attempt to fit the SL-TS2 model parameters $\alpha_{LL}$, $\bar{r}$, and $\xi$ to satisfy the "Boyer rules", setting $\alpha_L T_x \approx 0.165$, and $\alpha_G T_x \approx 0.055$. Then, the remaining parameters (activation energies) are determined based on the relaxation time master curve. For a more detailed discussion of SL-TS2, the physical meaning of the transition temperature $T_x$, and the non-divergence of the apparent activation energy at low temperatures, see Supporting Information.

To recap, all polymers (and possibly even some non-polymers, but this is a topic for a different study) that obey the "Boyer rules" can be described by a "universal" SL-TS2 free energy, with the same values of $\alpha_{LL}$, $\bar{r}$, and $\xi$. Furthermore, within the "corresponding states" framework, the activation energies are also universal, when normalized to the "energy scale" *($RT_x$)*, *R = 8.31 J/(mol K)* being the gas constant. (The proportionality of the liquid viscosity activation energy to $T_g$ has been discussed by several research groups – see, e.g., Bailly et al.[29] and Schmidtke et al.[55]) The specifics of an individual polymer are represented via its glass transition temperature, $T_g$, and fragility, *m*. One can then compute the "elementary time", $\tau_{el}$, and the normalized "solid-liquid transition temperature", $T_x/T_g$, as functions of *m*. The scaling rules described in Section 2.2 are then used to convert from the "master curve" to the relaxation time plots for specific polymers and to generate the corresponding VFTH or WLF equations. We can also use the "universal" SL-TS2 to study the dependence of the glass transition temperature on the cooling rate and to relate the experimental $T_g$ to the one obtained in molecular simulation. All of this is described in the next section.

## 3. Results and Discussion

### 3.1. Relaxation Times, WLF, and VFTH Equation Parameters

In Figure 1a, we plot the dielectric $\alpha$-relaxation data for the nine polymers described in Table 1. We apply the procedure described in Section 2.1 to collapse all the data onto one master curve (Figure 1b). The orange line is the equilibrium SL-TS2 relaxation time function, calculated with the model parameters summarized in Table 2.

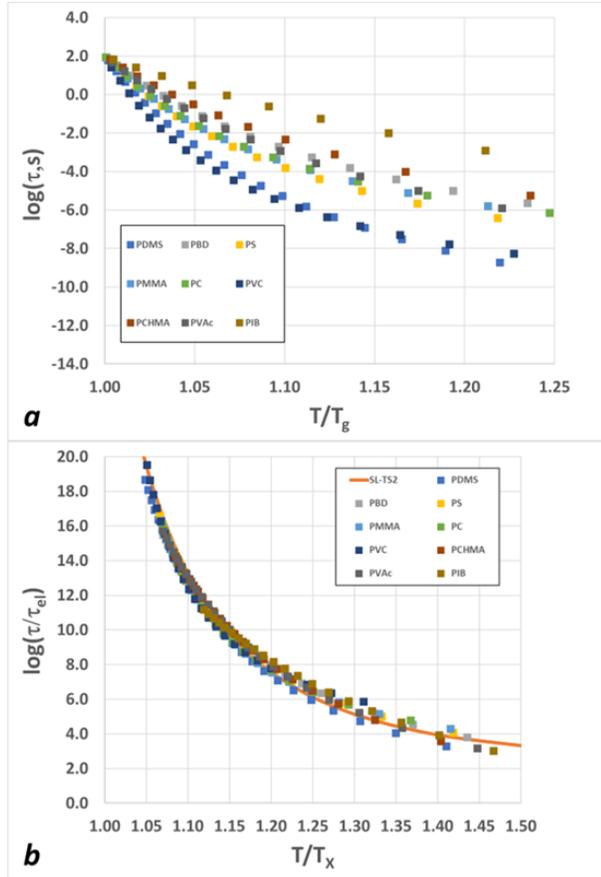

*Figure 1. (a) Logarithm of the α-relaxation time plotted as function of $T/T_g$ for the nine polymers described in Table 1. (b) Same data after the application of the transformation described in Section 2.1. The orange curve represents the SL-TS2 master curve, with model parameters given in Table 2.*

*Table 2. The universal SL-TS2 parameters for polymers*

| Parameter | Value |
|---|---|
| $\alpha_{LL}$ | 0.94 |
| $\bar{r}$ | 400 |
| $\xi$ | 0.03 |
| $E_1/RT_x$ | 8 |
| $E_2/RT_x$ | 134 |

(Interestingly, the size of the CRR, $\bar{r}$ = 400 ± 100, is similar to the value obtained by Krausser, Samwer, and Zaccone[21] for metallic glasses (~320). Their analysis was devoted to the description of shear

Page 15

transformation zones (STZ) that could be thought of as similar to CRR in the context of plastic deformation. It remains to be seen if this is more than coincidence.)

The shifting procedure of Section 2.1 relates the "universal" parameters to the specific "scale" parameters for individual polymers, i.e., $T_x$ and $\log(\tau_{el})$. The dependence of $\log(\tau_{el})$ and $T_x/T_g$ on the polymer fragility, *m*, is shown in Figures 2a and 2b, respectively. These dependencies can be approximated by analytical functions (see equations 4a and 4b). The parameters *k* and *b* in equation 5a-b – required for the calculations in equations 4a and 4b – are determined by fitting the SL-TS2 master curve and are given by: $k = 1.56$, and $b = 0.96$.

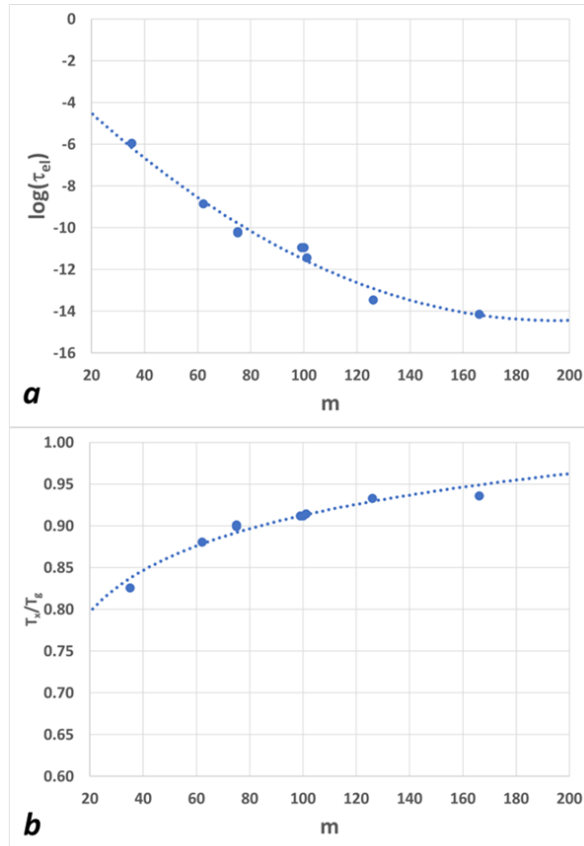

*Figure 2. (a) Fragility-dependent $\log(\tau_{el})$. Circles are from shifting procedure of Section 2.1 for the nine polymers of Table 1; the line is the analytical fit (equation 4a). (b) Ratio $T_x/T_g$ as a function of fragility m. Circles are from shifting procedure of Section 2.1 for the nine polymers of Table 1; the line is the analytical fit (equation 4a).*

The SL-TS2 solutions, for both the equilibrium case and the uniform cooling case with cooling rate Q = 1 K/min, are shown in Figure 3. The dependence of the specific volume on temperature (Figure 3a) agrees with the "Boyer rules"; the transition temperature at which the slope of the curve changes abruptly is equal to $T_g$ for Q ~ 1 K/min and depends on Q in a non-trivial fashion, as will be described in the next section.

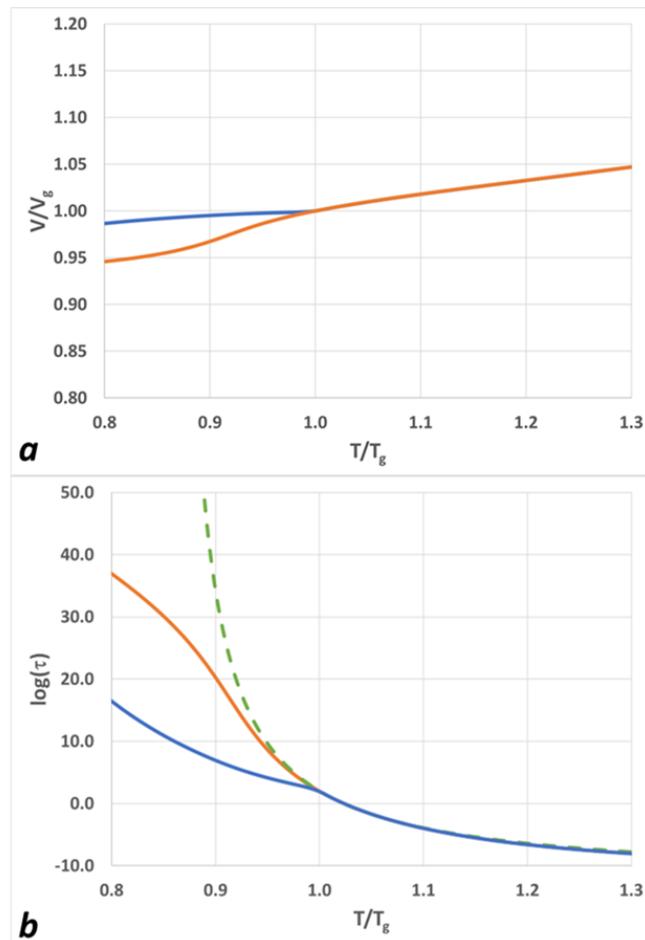

*Figure 3. (a) SL-TS2 specific volume vs. temperature. Orange – equilibrium; blue – cooling with rate Q = 1 K/min. (b) Logarithm of the relaxation time vs. temperature. Orange – equilibrium; blue – cooling with rate Q = 1 K/min; dashed green – WLF fit. The calculations are done for m = 100.*

The dependence of the relaxation time on the temperature (Figure 3b) is non-trivial. One can identify three regimes. The first regime ($T > 1.2T_g$) is the Arrhenius regime.[2,3,83,84] Within SL-TS2, it corresponds to the temperature region where $\psi \approx 0$, and the activation barrier is temperature-independent. The second regime ($0.9T_g < T < 1.2T_g$) is the VFTH regime, where the activation energy increases strongly as the temperature is decreased. Within the VFTH or WLF framework (often described based on the Adam-Gibbs[14] configurational entropy theory or the Doolittle[13] free-volume theory), activation energy diverges as temperature reaches the so-called Kauzmann[15,85] temperature $T_K \approx T_g - 50$ K. However, SL-TS2 anticipates that the activation energy remains finite even at low temperatures, consistent with several other models[16,18,20] and experiments.[22–24,58] Thus, in the third regime ($T < 0.9T_g$), there is an inflection point (coinciding with the thermodynamic transition temperature $T_X$) and transition to another Arrhenius behavior, albeit with significantly larger activation energy. Here, we concentrate only on the second (VFTH) regime. The approximate analytical fit to the numerical SL-TS2 master curve (the green curve in Figure 3b) is given by,

$$log\left(\frac{\tau_\alpha(T)}{\tau_{el}}\right) = \frac{k}{\left(\frac{T}{T_X}\right)-b} = \frac{kT_X}{T-bT_X} \tag{11}$$

where, as discussed above, $k = 1.56$, $b = 0.96$. Equation 11 can be written in the VFTH form as,

$$log(\tau_\alpha(T)) = log(\tau_{el}(m)) + \frac{D(m,T_g)}{T-T_0(m,T_g)} \tag{12}$$

where $D(m, T_g) = kT_g g(m)$, and $T_0(m, T_g) = bT_g g(m)$, where the function $g(m)$ is given by equation 5b. Alternatively, equation 11 can be written in the WLF form,

$$log\left(\frac{\tau_\alpha(T)}{\tau_g}\right) = -\frac{C_1(m)(T-T_g)}{C_2(m,T_g)+(T-T_g)} \tag{13a}$$

$$C_1(m) = \frac{kg(m)}{1-bg(m)} \tag{13b}$$

$$C_2(m, T_g) = T_g[1 - bg(m)] \tag{13c}$$

$$log(\tau_g) = 2 \tag{13d}$$

In addition, there are two interesting features that can be seen in Figure 3b. First, the inflection point (the onset of non-VFTH behavior at low temperatures and possible transition to the second Arrhenius region) corresponds to log(τ) ~ 15-20, or between 30 million and 3 trillion years. The lower bound (~30 million years) is comparable to the age of Dominican amber in well-known experiments of McKenna and co-workers,[22–24] in which they demonstrated the apparent transition to the low-temperature Arrhenius from the VFTH regime upon cooling deep below $T_g$. Of course, the agreement between model and experiment is not yet quantitative, and more analysis is needed. Second, we note that the Vogel or Kauzmann temperature, $T_0$, given by $T_0 = bT_x = bgT_g$, depends on the fragility $m$; the ratio $T_0/T_g$ varies from 0.75 (for $m$ = 20) to 0.91 (for $m$ = 200). The Adam-Gibbs[14] theory stipulated $T_0/T_g$ = 0.77, which agrees with our model in the limit of low-fragility ($m$ < 50) polymers.

In Figures 4 and 5, we compare the relaxation times for the nine polymers of Table 1 with the "universal model" of equations 12 or 13 (the two equations are, obviously, two different ways of writing the same function). Figure 4 depicts the lower-fragility ($m$ < 100) polymers, and Figure 5 shows the higher-fragility ($m$ ~ 100 – 180) polymers. The model parameters are summarized in Table 3.

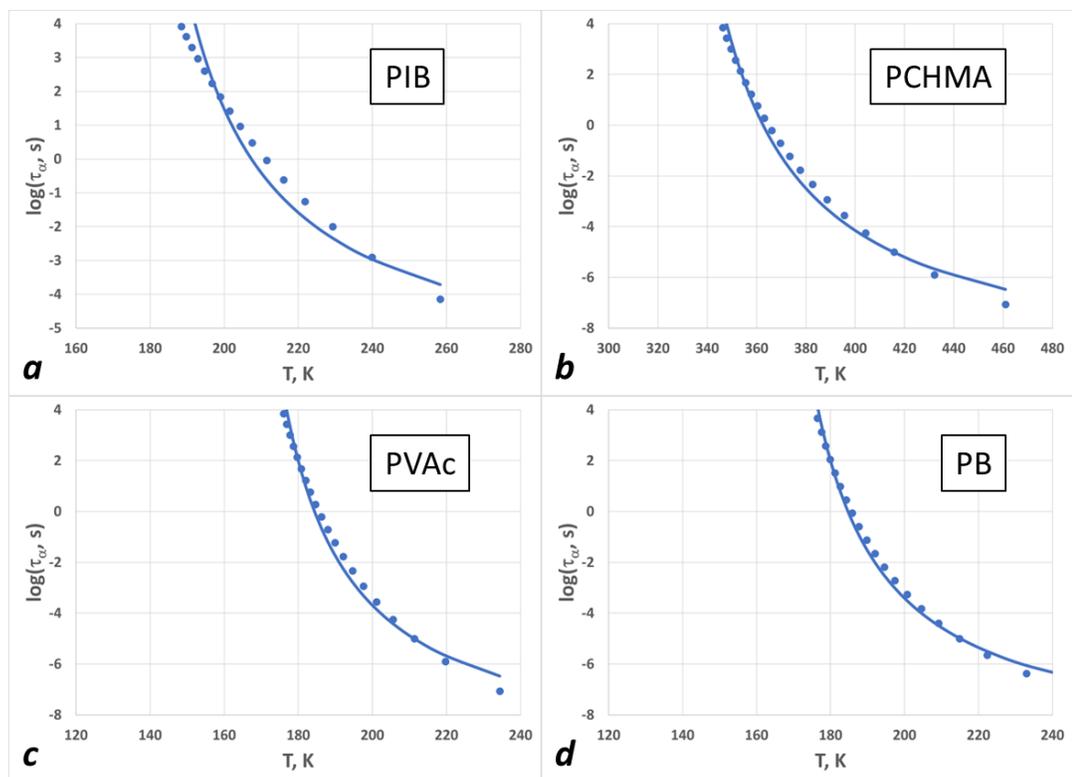

*Figure 4. The "universal model" fit vs. experimental data for the lower-fragility polymers: (a) PIB; (b) PCHMA; (c) PVAc; (d) PB.*

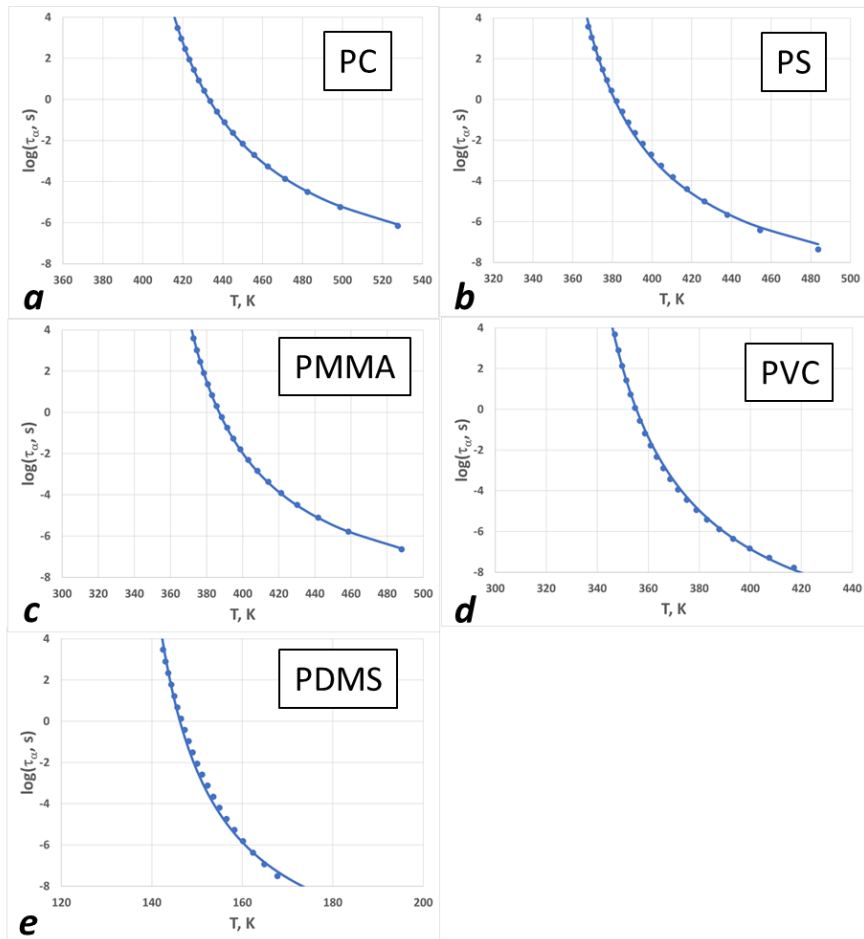

*Figure 5. The "universal model" fit vs. experimental data for the higher-fragility polymers: (a) PC; (b) PS; (c) PMMA; (d) PVC; (e) PDMS.*

*Table 3. Model parameters (see equations 3-5 and 14).*

| Polymer | $T_g$, K | $\log(\tau_{el})$ | $T_x/T_g$ | $C_1$ | $C_2$, K |
|---|---|---|---|---|---|
| PDMS | 144 | -12.84 | 0.94 | 14.8 | 14.2 |
| PBD | 180 | -9.46 | 0.91 | 11.5 | 22.4 |
| PIB | 198 | -6.66 | 0.88 | 8.7 | 31.3 |
| PVAc | 180 | -9.86 | 0.92 | 11.9 | 21.7 |
| PVC | 350 | -12.94 | 0.94 | 14.9 | 34.3 |
| PCHMA | 354 | -9.86 | 0.92 | 11.9 | 42.7 |
| PS | 373 | -10.61 | 0.92 | 12.6 | 42.6 |
| PMMA | 378 | -10.11 | 0.92 | 12.1 | 44.7 |
| PC | 423 | -9.99 | 0.92 | 12.0 | 50.5 |

The model curves fit the data reasonably well in the relaxation time range between $10^{-8}$ and $10^2$ s; one exception is the lowest-fragility polymer, PIB, where the fit quality is relatively poor. We will address this issue in the Discussion section. For now, we turn our attention to the ways one can use the new universal model to predict the relationship between the experimental and simulated $T_g$.

## 3.2. The Estimate of the Difference between the Simulated and Experimental Glass Transition Temperature

In a typical MD simulation aiming to calculate the glass transition temperature, a polymer is equilibrated at very high temperatures in the melt state. Then, the temperature is reduced by an increment $\Delta T$, and the system finds itself out of equilibrium. The temperature jump is then followed by equilibration over period $\Delta t$, at the end of which the specific volume, $V$, is computed. The process is then repeated multiple times. The rate of change, $\alpha = \Delta V/(V \Delta T)$, is the coefficient of thermal expansion. As the temperature is decreased, $\alpha$ remains nearly constant at first, but then decays sharply over a narrow temperature interval before settling at a new, lower, value. The mid-point of this transition interval is generally accepted as the glass transition temperature. This protocol is well-established and has been employed by many researchers over the years. It is also reasonably representative of the real-life dilatometry experiments; with one big caveat we now turn to.

In a recent comprehensive study by Afzal et al.,[44] $\Delta T$ = 20 K, and $\Delta t$ = 5 ns; other experimental studies have similar run parameters. This corresponds to a cooling rate $Q = |dT/dt| = \Delta T/ \Delta t$ = 4 K/ns or about $10^{11}$ K/min. In comparison, experimental cooling rates in DSC or dilatometry experiments are about $Q_0$ = 1 K/min. The gap of eleven orders of magnitude means that the computational $T_g$ is significantly higher than the experimental one, and this is universally acknowledged by all the researchers. The question is – can one predict the experimental $T_g$ given the computational one? Soldera et al.,[48,50,51] Sirk et al.,[45,46] and others proposed a formula based on the "universal" formulation of the WLF equation,[12] but acknowledged that it may not work for all polymers. To our knowledge, there have been few other attempts to develop a physics-based model relating the simulated and experimental $T_g$. Based on their study involving more than 300 amorphous polymers, Afzal et al.[44] proposed a phenomenological equation, *$T_g$(exp) = 0.77$T_g$(comp) + 21,* where *$T_g$(exp)* and *$T_g$(comp)* are, respectively, the experimental and computed glass transition temperatures, in K. The deviations between the predicted and measured $T_g$ vary from 0 to >50 K, and further understanding of the nature of these deviations is desired, as indeed was suggested by the authors. One question not answered in the paper was – how are the differences between the calculated and experimental $T_g$-s depend on the dynamic fragility of the polymer. Here we attempt to investigate this relationship in more detail.

The obvious question about the relationship between the simulation and experiment here is whether the simulation, with its characteristic time of a few nanoseconds, can capture enough of the relaxation processes that govern the real-life $T_g$. At temperatures close to $T_g$, the dielectric response of the material usually exhibits two or more peaks, with the lowest-frequency ("$\alpha$") peak centered around $\omega_\alpha \sim$ 0.01 s$^{-1}$ or $\tau_\alpha \sim$ 100 s. At higher temperatures, the first ("$\alpha$") and the second ("$\beta$") peak become closer and eventually merge at some temperature $T_{\alpha\beta}$, with the corresponding relaxation time $\tau_{\alpha\beta}$.[86-92] Between $T_g$ and $T_{\alpha\beta}$, the $\alpha$-relaxation time decays with temperature in accordance to the VFTH (or WLF) equation; above $T_{\alpha\beta}$, the decay is consistent with the Arrhenius-Andrade-Eyring model.[93] Empirically, $T_{\alpha\beta}$ -- roughly similar to the "Arrhenius temperature", $T_A$,[84] is on the order of 1.2 – 1.3 $T_g$, and $\tau_{\alpha\beta} \sim 10^{-5} - 10^{-4}$ s. Thus, the experimental studies of the glass transition probe the WLF behavior, and the simulations are restricted to the Arrhenius region. It is thus necessary to ensure that the relaxation time description transitions smoothly from VFTH to Arrhenius – or the VFTH equation itself provides a reasonably good fit for at least a portion of the Arrhenius region (see Supporting Information).

The scaling analysis described above suggests another interesting question and challenge. The "elementary time", $\tau_{el}$, describes the "CRR equilibration time at infinite temperature". This implies that the time of at least $\tau_{el}$ is needed to equilibrate a CRR even in a liquid state, at temperatures way above the real-life glass transition temperature. If the simulation time, $\Delta t$, is significantly less than $\tau_{el}$, the equilibration is "incomplete". It is difficult to say what it means physically; however, we can conceivably expect a dramatic change as one transitions from a regime where $\Delta t \ll \tau_{el}$ to the one where $\Delta t \gg \tau_{el}$. Given that in simulations, $\Delta t \sim 1 - 10$ ns, this change is expected to occur for polymers with m ~ 60-80.

Based on the Frenkel-Kobeko-Reiner (FKR) hypothesis[94–97] and simple dimensional analysis, one can relate the heating rate to the corresponding relaxation time (or frequency) as $\tau_g(Q)Q/T_g(Q) \approx const$. This can be re-written as,

$$log\left(\frac{T_g(Q)}{T_g}\right) = log\left(\frac{Q}{Q_0}\right) + log\left(\frac{\tau(T_g(Q))}{\tau_g}\right) \tag{14}$$

Here, $T_g$ is the experimental glass transition temperature, $Q_0 \sim 1$ K/min is the real-life experiment cooling rate, and $\tau_g = 10^2$ s is the α-relaxation time at $T = T_g$. Similarly, $Q$ is the cooling rate in computer simulations, $T_g(Q)$ is the transition temperature in the simulation, and $\tau(T_g(Q))$ is the α-relaxation time at $T = T_g(Q)$. As discussed above,

$$log\left(\tau(T_g(Q))\right) = log(\tau_{el}) + \mathbb{F}\left(\frac{T_g(Q)}{T_x}\right) \tag{15}$$

Furthermore, if we denote $x = \frac{T_g(Q)}{T_g}$, then we can easily obtain,

$$log(x) = log\left(\frac{Q}{Q_0}\right) + \mathbb{F}\left(\frac{x}{g(m)}\right) - \mathbb{F}\left(\frac{1}{g(m)}\right) \tag{16a}$$

Or

$$log\left(\frac{Q}{Q_0}\right) = log(x) - \mathbb{F}\left(\frac{x}{g(m)}\right) + \mathbb{F}\left(\frac{1}{g(m)}\right) \tag{16b}$$

This provides an implicit formula for x as a function of $Q/Q_0$. In Figure 6a, we plot the solutions of this equation for several values of fragility, m, between 60 and 180; Figure 6b depicts the dependence of $\frac{T_g(Q)}{T_g}$ on m for $log\left(\frac{Q}{Q_0}\right) = 11$.

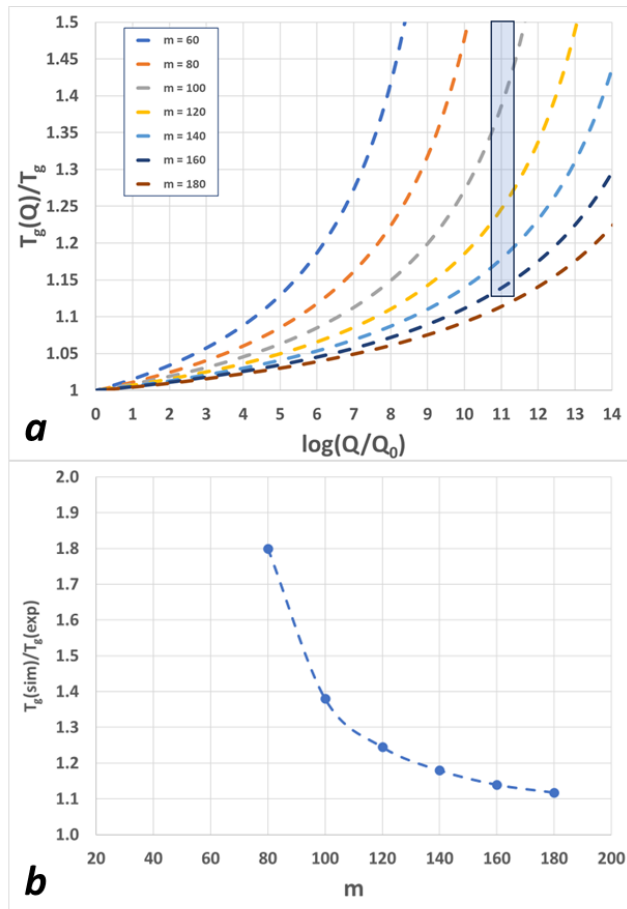

Figure 6. Calculated dependence of $T_g$ on the cooling rate Q for polymers with different fragilities between 60 and 180. The light-blue rectangle corresponds to approximate conditions of a typical atomistic MD simulation, like that of Afzal et al.[44]

The formula of equation 16b is numerically identical to the Soldera approach,[50] which is based on the WLF equation, provided that the WLF coefficients $C_1$ and $C_2$ are given by equations 13b and 13c. It is important to note that there are two sources of uncertainty as one attempts to relate the simulated $T_g$ to the real-life one. First, the "effective cooling rate" of the computer experiment may not be exactly given by $\Delta T/\Delta t$ – this is only an order of magnitude estimate. As one can see from Figure 6a—6b, the dependence of $T_g(Q)$ on $Q$ is much steeper in the high-Q region than in the lower-Q one, so even a small uncertainty in Q translates into a large error in $T_g$. Second, since the simulations cannot reach the experimental relaxation times, it is not clear whether the force field is "properly tuned" to predict the correct fragility. One could assume that a "good" forcefield (like OPLS) should correctly capture the differences between the "strong" and "fragile" materials – but probably not the absolute magnitude of fragility. (One should note that even based on experimental analysis, fragility for a given polymer is determined within ±20.)[98,99] We thus attempt to re-analyze the Afzal simulation results for the same nine polymers of Table 1, allowing for the fragility variation of ±20 from the best-fit value. The results of this analysis are shown in Table 4 and Figure 7.

Table 4. Simulated (data from Afzal et al.[44]), experimental, and predicted glass transition temperatures for the nine polymers of Table 1. See text for details.

| Polymer | $T_{g\_calc}$, K | $T_{g\_exp}$, K | R | $m_1$ | $R_1$ | $m_2$ | $R_2$ | $T_{min}$, K | $T_{max}$, K |
|---|---|---|---|---|---|---|---|---|---|
| PMMA | 455 | 378 | 1.20 | 100 | 1.4 | 140 | 1.16 | 325 | 392 |
| PS | 438.5 | 367 | 1.19 | 100 | 1.4 | 140 | 1.16 | 313 | 378 |
| PDMS | 193.3 | 144 | 1.34 | 110 | 1.3 | 150 | 1.14 | 149 | 170 |
| PBD | 286.5 | 180 | 1.59 | 80 | 2 | 120 | 1.25 | 143 | 229 |
| PVC | 423.2 | 350 | 1.21 | 130 | 1.24 | 170 | 1.12 | 341 | 378 |
| PC | 607 | 423 | 1.43 | 100 | 1.4 | 140 | 1.16 | 434 | 523 |
| PVAc | 406 | 307 | 1.32 | 80 | 2 | 120 | 1.25 | 203 | 325 |
| PCHMA | 461 | 354 | 1.30 | 80 | 2 | 120 | 1.25 | 231 | 369 |
| PIB | 198.6 | 198 | 1.00 | No Prediction Possible | | | | | |

In Table 4, the second column shows the simulation results of Afzal et al.[44] and the third column contains experimental values (same as in Table 1); the fourth column is the ratio of the two. The fifth column contains the lower-bound for the fragility, and the sixth column is the estimate for the corresponding (upper-bound) ratio of R = $T_g(sim)/T_g(exp)$. Likewise, the seventh column contains the upper-bound for the fragility, and the eighth column is the lower-bound for the ratio R. Finally, the ninth and tenth columns are the lower and upper bounds for the actual $T_g$ based on the simulation value and the ratios. Note that for PIB, the upper bound for the fragility is still less than 80, and thus no prediction is possible within our framework. We will discuss this topic further in the Discussion section.

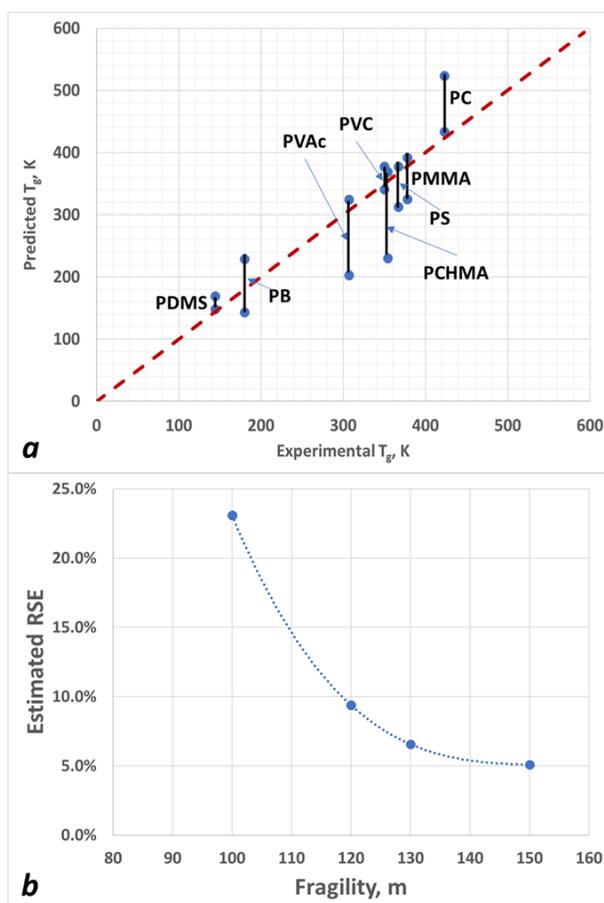

*Figure 7. (a) Predicted $T_g$ vs. experimental one. The blue circles are upper (the tenth column of Table 3 vs. the second column), and lower (the ninth column vs. the second column) bounds. The dashed red line is the Y = X line. The labels show which polymer is represented by each specific circle. (b) Estimated relative standard error (RSE) for the predicted $T_g$ as a function of the polymer fragility.*

Figure 7a depicts the relationship between the experimental and predicted $T_g$. From this figure, we can draw two important conclusions. First, the proposed approach enables one to estimate experimental ("slow cooling rate") $T_g$ from the simulated ("fast cooling rate") one. This is consistent with the work of Soldera et al.[48,50,51] and is encouraging. Second, the precision of this estimate depends strongly on the fragility of the polymer. For the higher-fragility ones (PDMS and PVC), the experimental $T_g$ is estimated within ±10 K. For the intermediate-fragility polymers (PC, PMMA, PB, and PS), the error increases to ±40 K, and for the lower-fragility ones (PCHMA and PVAc) to ±50 K. We visualize this dependence in Figure 7b, showing the estimated relative standard error (RSE) for the predicted $T_g$ as a function of the polymer fragility. The estimate is based on the assumption that the fragility itself has a standard error of ±20. Thus, the accuracy and precision of the prediction are expected to be very good for the high-fragility polymers (m > 130) and worsen dramatically as the fragility drops below 100. As discussed above, the model fails to make a prediction for the "strong" polymer PIB, as its CRR elementary time $\tau_{el}$ is predicted to be significantly larger than the nanosecond timescale probed in the simulations.

In practice, when simulation is performed for a new polymer, its fragility is unknown. Thus, to estimate the experimental $T_g$ based on the simulated one, it is necessary to assume what the fragility is. Based on the Afzal et al.[44] analysis, it would seem that the best choice is to pick *m ≈ 110*. This would correspond to the ratio $T_g$(exp)/$T_g$(sim) ≈ 0.77, in agreement with the observed average value.[44]

### *3.3. Discussion*

In this paper, we formulate two main hypotheses. First, the model outlined here suggests that the VFTH region of the relaxation time vs. temperature curve is "universal" for many polymers when plotted in non-dimensional form: $log\left(\frac{\tau_\alpha(T)}{\tau_{el}(m)}\right) = \mathbb{F}\left(\frac{T}{T_g g(m)}\right)$. This is, in some way, an extension of earlier ideas of Williams, Landel, and Ferry,[12] Ding and Sokolov,[28] Wang and Porter,[27] Bailly et al.,[29] van Krevelen,[2] and Bicerano.[3] In this framework, the elementary time, $\tau_{el}$, and the glass transition temperature, $T_g$, set the time and temperature scales; for convenience, we use $T_g$ and fragility and express the elementary time as a function of fragility. This scaling and universality is suggested naturally by the Boyer-Spencer and Simha-Boyer scaling for the thermal expansion, combined with the idea that the relaxation time is related to the free volume.[27,29] We use the SL-TS2 theory to regress the universal parameters, but other free-volume approaches could be utilized as well. The SL-TS2 master curve was parameterized using VFTH and WLF functional form, and the WLF parameters $C_1$ and $C_2$ have been regressed as function of $T_g$ and $m$.

Second, the scaling and universality derived here can be used to describe the relationship between the atomistic Molecular Dynamics simulated glass transition temperature and the experimental one. We analyze the simulation results of Afzal et al.[44] and estimate the accuracy and precision of the predicted $T_g$ based on the simulated one. Crucially, we find that the precision of the prediction depends on the polymer fragility – the predictions are expected to be more precise for high-fragility polymers and less precise for lower-fragility one. (For the case of the lowest-fragility polymer in our study, poly(isobutylene), we find it impossible to make a prediction for reasons discussed below).

The scaling analysis described above has one profound implication. The "elementary time", $\tau_{el}$, describes the "CRR equilibration time at infinite temperature". This implies that the time of at least $\tau_{el}$ is needed to equilibrate a CRR even in a liquid state, way above the real-life glass transition. If the simulation or measurement time, $\Delta t$, is significantly less than $\tau_{el}$, the equilibration is "incomplete". It is difficult to say what it means physically; however, we can conceivably expect a dramatic change as one transitions from a regime where $\Delta t \ll \tau_{el}$ to the one where $\Delta t \gg \tau_{el}$. For broadband dielectric spectroscopy (~$10^{-7} - 10^{1}$ s), this change is observed only for lowest-fragility polymers, which is why the model fails to correctly describe the relaxation behavior of PIB whose $\tau_{el}$ is predicted to be on the order of $10^{-6}$ s. For NMR experiments ($\sim 10^{-10} - 10^{-8}$ s) and for atomistic Molecular Dynamics simulations ($\sim 10^{-11} - 10^{-10}$ s), this transition is expected to take place for intermediate-fragility polymers, leading to decreased precision in the prediction of experimental $T_g$ from simulated ones.

It is important to realize that the proposed picture is somewhat in odds with the standard Angell classification of "strong" and "fragile" glass-formers.[100–106] In Angell's view, the "strong" materials are physically different from the "fragile" ones. Thus, per Angell, the WLF parameter $C_1$ should be about 16-17, independent of the material fragility.[103] In our view, the "strong" and the "fragile" materials simply occupy different regions of the master curve. The WLF parameter $C_1$ is indeed about 16-17 for high-fragility polymers, but decreases significantly for the lower-fragility ones. This idea is consistent with the concept of "vertical shift" proposed, e.g., by Bailly et al.[29]

Obviously, this is only an initial hypothesis and more studies are needed to validate or disprove it. It would be interesting to see whether the scaling relationship derived for the "Boyer polymers" is valid for those polymers that have strong deviations from the Boyer-Spencer and Simha-Boyer rules. Other topics to consider include the dependence of the model parameters on molecular weight within the homologous series, as well as the applicability of the proposed scaling and universality to non-polymeric glass-formers. These will be topics of future studies.

## 4. Conclusions

We developed a new scaling model that allows for a superposition of various relaxation time vs. temperature curves onto a single master curve. Thus, a relaxation time dependence of a polymeric glass-former is fully determined by only two parameters – the glass transition temperature, $T_g$, and the fragility, $m$. The master curve, combined with the Boyer-Spencer and Simha-Boyer rules for thermal expansion, is captured by a numerical solution resulting from the SL-TS2 free energy minimization. The master curve can also be parameterized by a VFTH-like function, and thus, one can obtain analytical expressions for the dependence of the WLF parameters $C_1$ and $C_2$ on $T_g$ and $m$.

The model also allows one to compute the relationship between experimental and simulated glass transition temperature. We have shown that this relationship is generally quite accurate, although its precision depends strongly on the polymer fragility; in the limit of very high fragility ($m$ ~ 140 – 160), the prediction is thus both accurate and precise. The findings will be useful in developing novel simulation techniques and workflows.

## Associated Content

*Supporting Information*

The Supporting Information is available free of charge at…

Equations of the Sanchez-Lacombe Two-State, Two-(Time)Scale Theory – Equilibrium; Equations of the Sanchez-Lacombe Two-State, Two-(Time)Scale Theory – Dynamics; Equations of the Sanchez-Lacombe Two-State, Two-(Time)Scale Theory – Casalini-Roland Dynamic Scaling; Numerical Procedures for the Calculation of $\nu[T]$ and $\psi[T]$; The Relationship between the SL-TS2 and Arrhenius Functions for the $\alpha$-Relaxation Time.

## Author Information

*Corresponding Author*

Valeriy V. Ginzburg – *Department of Chemical Engineering and Materials Science, Michigan State University, East Lansing, Michigan 48824-1312, United States;* orcid.org/0000-0002-2775-5492; Email: vvg851966@gmail.com

*Authors*

Oleg V. Gendelman – *Faculty of Mechanical Engineering, Technion, Haifa 32000, Israel;* orcid.org/0000-0002-4750-2708

Riccardo Casalini -- *Chemistry Division, Naval Research Laboratory, 4555 Overlook Avenue SW, Washington, D.C. 20375, USA;* orcid.org/0000-0002-5717-4103

Alessio Zaccone – *Department of Physics, University of Milan, 20122 Milan, Italy;* orcid.org/0000-0002-6673-7043

## Acknowledgments

OVG is very grateful to Israel Science Foundation (ISF) for financial support, grant 2598/21. VVG thanks Dr. Jozef Bicerano for helpful suggestions.

## Abbreviations

DoF – Degree of Freedom; ECNLE – Elastically Cooperative Nonlinear Langevin Equation (theory); EoS – Equation of State; LCL-CFV – Locally Correlated Lattice/Cooperative Free Volume (theory); NMR – Nuclear Magnetic Resonance; SL-TS2 – Sanchez-Lacombe/Two-State, Two (time)Scale (theory); VFTH – Vogel-Fulcher-Tammann-Hesse (equation); WLF – Williams-Landel-Ferry (equation);

Supporting Information for:

Universality of Polymer Dynamics near the Glass Transition and the Relationship between the Simulated and Experimental Glass Transition Temperatures of Amorphous Polymers

Valeriy V. Ginzburg[1,*], Oleg V. Gendelman[2], Riccardo Casalini[3], and Alessio Zaccone[4]

[1]Department of Chemical Engineering and Materials Science, Michigan State University, East Lansing, Michigan, USA

[2]Faculty of Mechanical Engineering, Technion, Haifa, Israel

[3]Chemistry Division, Naval Research Laboratory, 4555 Overlook Avenue SW, Washington, D.C., USA

[4]University of Milan, Department of Physics, via Celoria 16, 20133 Milano, Italy

# 1. Equations of the Sanchez-Lacombe Two-State, Two-(Time)Scale Theory-- Equilibrium

The volume fractions of different species are given by,

$$\phi_S = v \frac{\psi r_S}{r} \tag{S1a}$$

$$\phi_L = v \frac{[1-\psi] r_L}{r} \tag{S1b}$$

$$\phi_V = 1 - v \tag{S1c}$$

After defining $\varepsilon^* = \frac{2 k_B T}{z \varepsilon_{SS}}$ and $\tilde{T} = \frac{k_B T}{\varepsilon^*}$, eq 12 of the main text is transformed into,

$$G \equiv \frac{G}{\varepsilon^*}\left(\frac{v_0}{V_{CRR}}\right) = T\left[\psi \ln \phi_S + (1-\psi)\ln \phi_L + r\left\{\frac{1}{v} - 1\right\}\ln \phi_V\right]$$

$$-\frac{r}{v}\left[\phi_S^2 + 2\alpha_{SL}\phi_S\phi_L + \alpha_{LL}\phi_L^2\right] \tag{S2}$$

Here, we defined $\alpha_{SL} = \varepsilon_{SL}/\varepsilon_{SS}$ and $\alpha_{LL} = \varepsilon_{LL}/\varepsilon_{SS}$; following a traditional Berthelot geometric rule, we postulate that $\alpha_{SL} = \sqrt{\alpha_{LL}}$.

Differentiating eq S2 with respect to $\psi$ and $v$, we obtain,

$$\ln \frac{\psi}{1-\psi} + \Delta \tilde{S} - \frac{\Delta \tilde{U}}{\tilde{T}} = 0 \tag{S3a}$$

$$\tilde{T}\left[\ln(1-v) + v\left\{1 - \frac{1}{r}\right\}\right] + v^2 J = 0 \tag{S3b}$$

Here,

$$\Delta S = \ln\frac{r_S}{r_L} + \frac{r_L - r_S}{r} + (r_L - r_S)\left(\frac{1}{v} - 1\right)\ln\frac{1}{1-v} \quad \text{(S4a)}$$

$$-\Delta U = \frac{v}{r^2}\left\{(r_L - r_S)\psi^2 r_S^2 - 2\psi r_S^2 r_L\right\}$$

$$+2\alpha_{LS}\frac{v}{r^2}\left\{(r_L - r_S)\psi(1-\psi)r_S r_L - r_S r_L\left[(1-\psi)r_L - \psi r_S\right]\right\} \quad \text{(S4b)}$$

$$+\alpha_{LL}\frac{v}{r^2}\left\{(r_L - r_S)(1-\psi)^2 r_L^2 + 2(1-\psi)r_L^2 r_S\right\}$$

$$J = \left(\frac{\psi r_S}{r}\right)^2 + 2\alpha_{LS}\left(\frac{\psi r_S}{r}\right)\left(\frac{\{1-\psi\}r_L}{r}\right) + \alpha_{LL}\left(\frac{\{1-\psi\}r_L}{r}\right)^2 \quad \text{(S4c)}$$

In the following, we will use parameters $\bar{r} = 0.5(r_S + r_L)$ and $\xi = (r_L - r_S)/(r_L + r_S)$, rather than $r_L$ and $r_S$. The newly defined parameter $\xi$ equals to one-half of the relative volume difference between the "pure" liquid and "pure" solid states.

For any given $\tilde{T}$, the equilibrium values of $\psi$ and $v$ are the solutions of eqs S3a—S3b and denoted $\psi_{eq}[\tilde{T}]$ and $v_{eq}[\tilde{T}]$. The equilibrium CRR volume and specific volume are given by,

$$V_{CRR} = v_0 \frac{r}{v} \quad \text{(S5a)}$$

$$V_{sp} = V_0 \frac{r}{r_S v} \quad \text{(S5b)}$$

$$r = \psi r_S + (1 - \psi) r_L \quad \text{(S5c)}$$

Here, $V_0$ is the "ideal" specific volume of the material at T = 0 K (pure solid state, no voids).

## 2. Equations of the Sanchez-Lacombe Two-State, Two-(Time)Scale Theory – Relaxation Times

The $\beta$- (Johari-Goldstein[1,2]) and $\alpha$-relaxation times are given by,

$$\tau_\beta[\tilde{T}] = \tau_{el}\, exp\left[\frac{\widetilde{E_1}}{\tilde{T}}\right] \tag{S6a}$$

$$\tau_\alpha[\tilde{T}] = \tau_{el}\, exp\left[\frac{\widetilde{E_1}}{\tilde{T}} + \frac{\widetilde{E_2}-\widetilde{E_1}}{\tilde{T}}\psi[\tilde{T}]\right] \tag{S6b}$$

The dimensionless activation energies $E_1$ and $E_2$ are defined as,

$$\widetilde{E_1} = \frac{E_1}{\varepsilon^*} \tag{S7a}$$

$$\widetilde{E_2} = \frac{E_2}{\varepsilon^*} \tag{S7b}$$

Here, $E_1$ and $E_2$ are the "standard" Arrhenius activation energies of the liquid and solid states, respectively. For more details, see Ginzburg et al.[3–5]

## 3. Equations of the Sanchez-Lacombe Two-State, Two-(Time)Scale Theory – Casalini-Roland Dynamic Scaling

As already discussed,[4,5] within SL-TS2, the "τTV" Casalini-Roland scaling is automatically satisfied in equilibrium, provided that,

$$V_0[P] = V_0[0]\, exp\left(-\frac{P}{P_v}\right) \tag{S8a}$$

$$E_i[P] = E_i[0]\, exp\left(\frac{P}{P_T}\right) \tag{S8b}$$

$$\frac{P_v}{P_T} = \gamma \tag{S8c}$$

Here, $E_i$ refers to all model parameters with the units of energy or temperature, i.e., $\varepsilon^*$, $E_1$, and $E_2$. In the *(T,P)* plane, all points with $\tilde{T} \equiv \left[\frac{T}{\varepsilon^*[P]}\right] = const.$, are isochronal, at least at relatively small pressures where the Casalini-Roland scaling holds. The zero-temperature specific volume, $V_0$, depends on the pressure, as shown in eq S8a. Then, condition (S8c) ensures that the Casalini-Roland relationship holds (for more details, see our earlier papers).[4,5]

# 4. Numerical Procedures for the Calculation of $v[T]$ and $\psi[T]$.

## *4.1. Equilibrium Solution*

The flowchart for the calculation of the equilibrium SL-TS2 functions is given in Figure S1. We start by selecting $\psi$ as the independent variable and generating an array $\Psi_i$ (with the first element $\Psi_1$ = 0.01, the last element $\Psi_{99}$ = 0.99, the increment being 0.01). The free energy minimization equations S3a and S3b are re-written in a form,

$$\tilde{T} = \frac{\Delta \tilde{U}}{ln\frac{\psi}{1-\psi}+\Delta \tilde{S}} \tag{S9a}$$

$$\tilde{T} = -\frac{v^2 J}{\left[ln(1-v)+v\left\{1-\frac{1}{r}\right\}\right]} \tag{S9b}$$

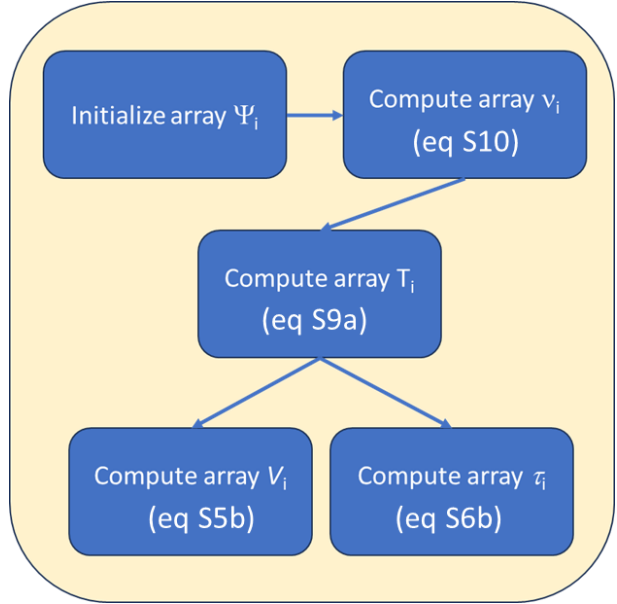

*Figure S 1. Flowchart for the calculation of equilibrium SL-TS2 solutions.*

Eliminating $T$ from eqs S9a-b and rearranging the terms, we obtain,

$$ln\frac{1}{1-v} = -\frac{v^2 J}{\Delta \tilde{U}}\left[\Delta \tilde{S} + ln\frac{\psi}{1-\psi}\right] + v\left\{1 - \frac{1}{r}\right\} \quad (S10)$$

We solve eq S10 iteratively, using the Picard mixing rule. Once the function $v_{eq}[\psi_{eq}]$ (represented as array $v_i$) is computed, we use eq S9a to calculate the corresponding temperature array, $\tilde{T}_i$. Next, the arrays corresponding to specific volume, $V_i$ and relaxation time, $\tau_i$, are calculated using eqs S5b and S6b below. The calculation results are shown in Figures S2 and S3, and the SL-TS2 parameters used for the calculation are summarized in Table S1 (same as Table 2 in the main text). Note that the transition temperature, $T_x$, is estimated to be equal to 0.3396T* (where T* = $\varepsilon$*/$k_B$).

Table S 1. Universal dimensionless SL-TS2 parameters

| Parameter | Value |
|---|---|
| $\alpha_{LL}$ | 0.94 |
| $\bar{r}$ | 400 |
| $\xi$ | 0.03 |
| $E_1/RT_x$ | 8 |
| $E_2/RT_x$ | 134 |

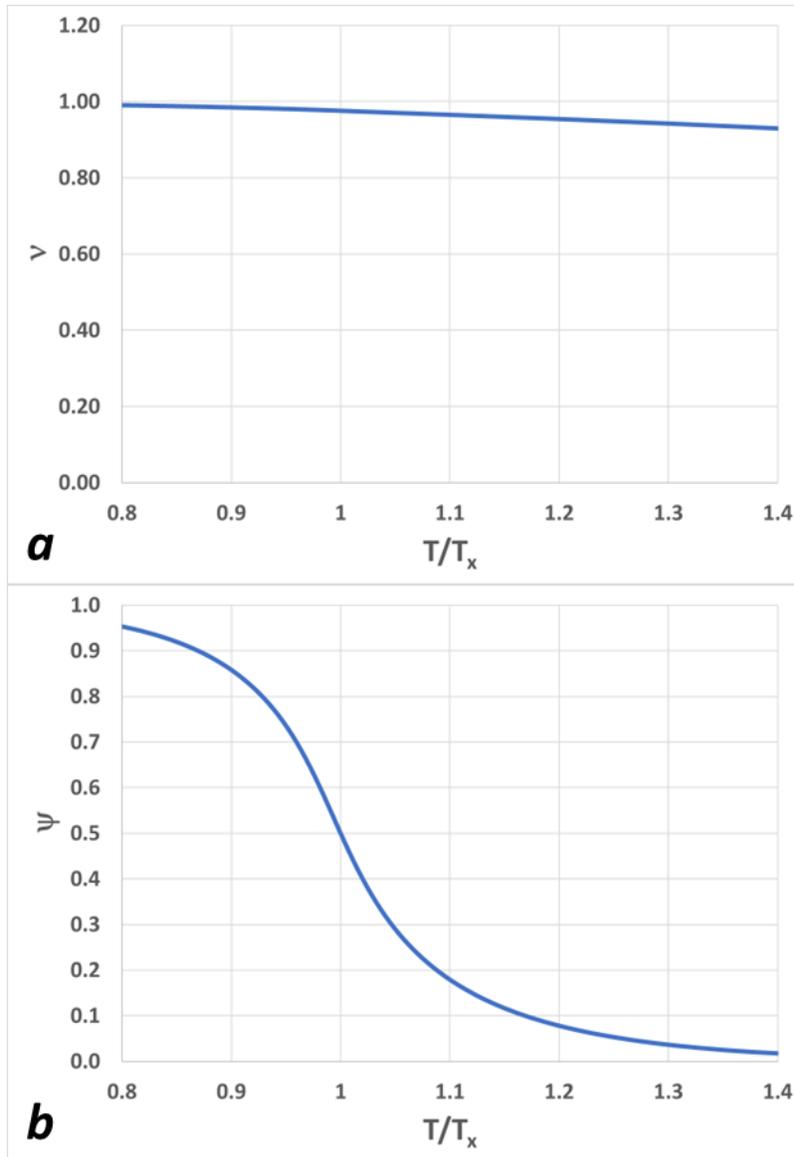

Figure S 2. The equilibrium SL-TS2 variables: (a) $v$ (occupancy) and (b) $\psi$ (solid fraction) as a function of reduced temperature. See text for more details.

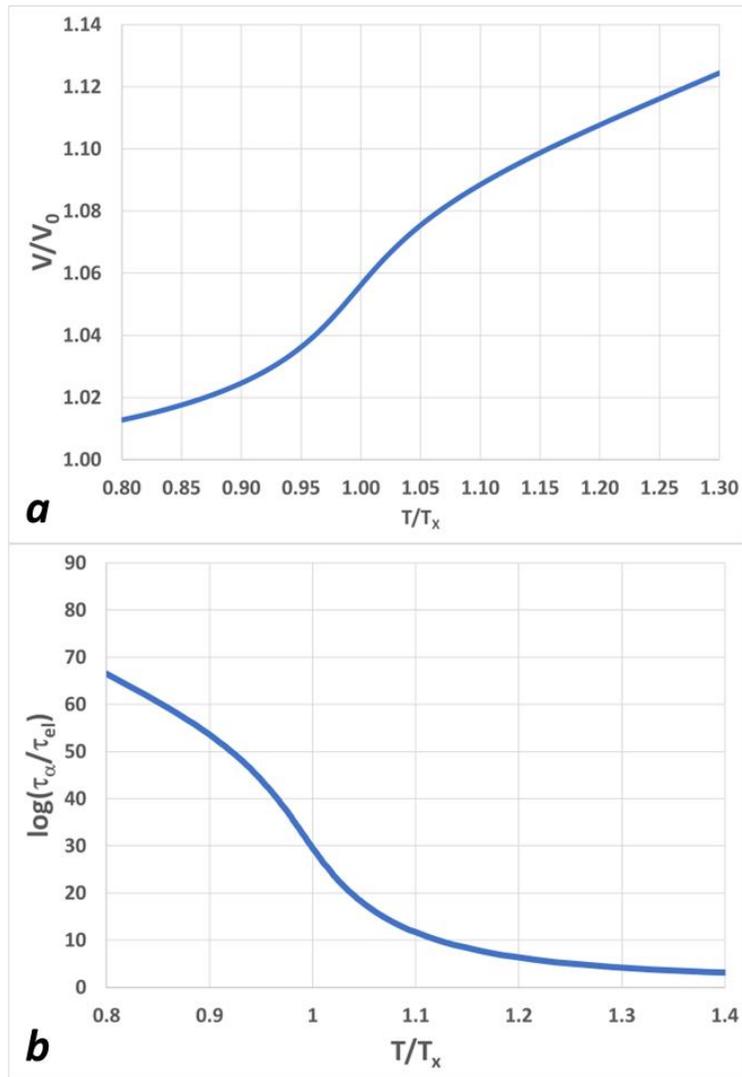

*Figure S 3. The equilibrium SL-TS2 variables: (a) specific volume (normalized to its equilibrium value at T → 0) and (b) α-relaxation time (normalized to $\tau_{el}$). See text for more details.*

### 4.2. Time-Dependent Solution

Let us now consider an experiment where the material is cooled from its equilibrium liquid state ("melt") with a constant cooling rate Q = |dT/dt|, and the relaxation time and specific volume are

measured as a function of temperature. The flowchart for determining the time-dependent solutions to dynamic SL-TS2 equations is given in Figure S4.

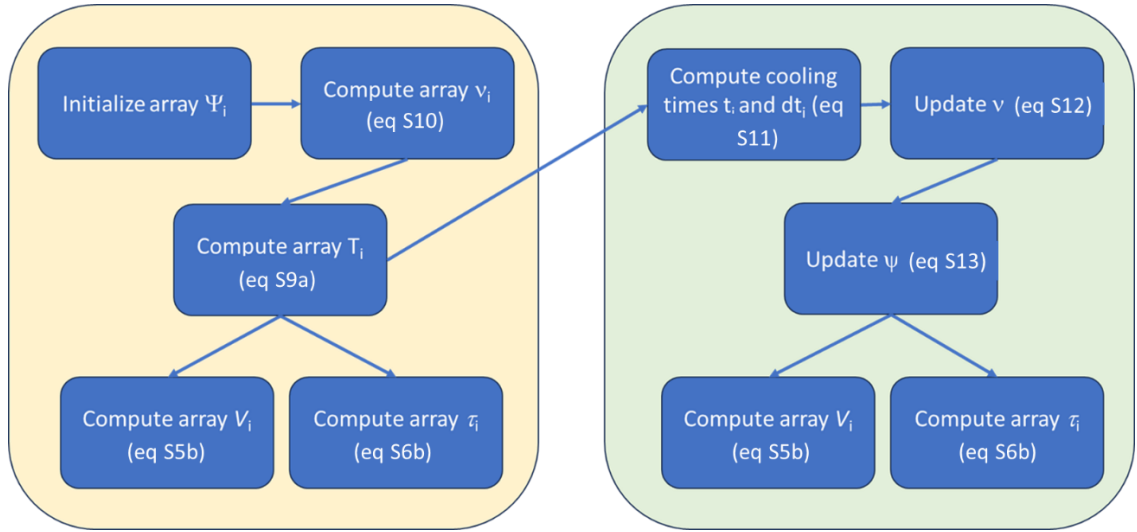

*Figure S 4. Flowchart for solving the dynamic SL-TS2 equations under conditions of simulated cooling with constant cooling rate Q.*

The left panel of Figure S4 describes the equilibrium calculations, discussed in the previous section. For each temperature, $T_i$, in our array of temperatures, we can determine the time, $t_i$, and the "residence time", $dt_i$,

$$t_i = \frac{T_1 - T_i}{Q} \qquad (S11a)$$

$$dt_i = \frac{T_{i+1} - T_i}{Q} \qquad (S11b)$$

In our earlier paper,[5] we made the following approximation,

$$v_i = \begin{bmatrix} v_{eq}[T_i], dt_i > \tau_{\alpha,i} \\ v_{eq}[T_g], dt_i \leq \tau_{\alpha,i} \end{bmatrix}$$

with the idea that the glass transition temperature is the point where the residence time becomes equal to the equilibrium relaxation time. This can be considered a zero-order approximation to the true solution of eqs 8a-b of the main text. Here, we propose to use a first-order approximation,

$$v_i = \lambda_i v_{i-1} + (1 - \lambda_i) v_{eq,i} \tag{S12a}$$

$$\lambda_i = exp\left[-\frac{dt_i}{\tau_{eq}[T_i]}\right] \tag{S12b}$$

Then, we can update the solid fraction, $\psi_i$, by iteratively solving the following equation,

$$\ln\frac{\psi_i}{1-\psi_i} = \frac{\Delta \tilde{U}}{v_i^2 J}\left[\ln[1 - v_i] + v_i\left\{1 - \frac{1}{r}\right\}\right] - \Delta \tilde{S} \tag{S13}$$

Next, we compute the effective specific volume and relaxation time arrays, using eqs S5 and S6b and substituting $\psi_i$ and $v_i$ instead of $\psi_{eq}$ and $v_{eq}$.

So far, we describe everything in terms of $\varepsilon^*/R$ (for the temperature scale), $V_0$ (for the specific volume scale), and $\tau_\infty$ (for the timescale); however, experimentally we only know $T_g$, $V_g$, and $\tau_g$. How do we translate from one scale to another?

Let us consider, yet again, the equilibrium solution. As already mentioned above, the point $\psi = 0.5$ (the free energies of the solid and liquid states are equal) corresponds to reduced temperature $\tilde{T} \approx 0.3396$. We now define this point as $\tilde{T}_x$ (or $T_x$ in real temperature units). This is the same $T_x$ we used in Section 2 of the main text, and, as already mentioned, the red curve in Figure 1b of the main text corresponds to substituting the equilibrium solution for $\psi$ into eq S6b. Thus, $\frac{T}{T_x} = \frac{\tilde{T}}{\tilde{T}_x} = \frac{\tilde{T}}{0.3396}$. We now apply eqs 4a—4b of the main text to convert $T/T_x$ to $T/T_g$. Based on this approach, we can plot, e.g., the equilibrium and non-equilibrium values of the internal variables $\psi$ and $v$, assuming fragility *m = 100* and cooling rate *Q = 1 K/min* (Figure S5).

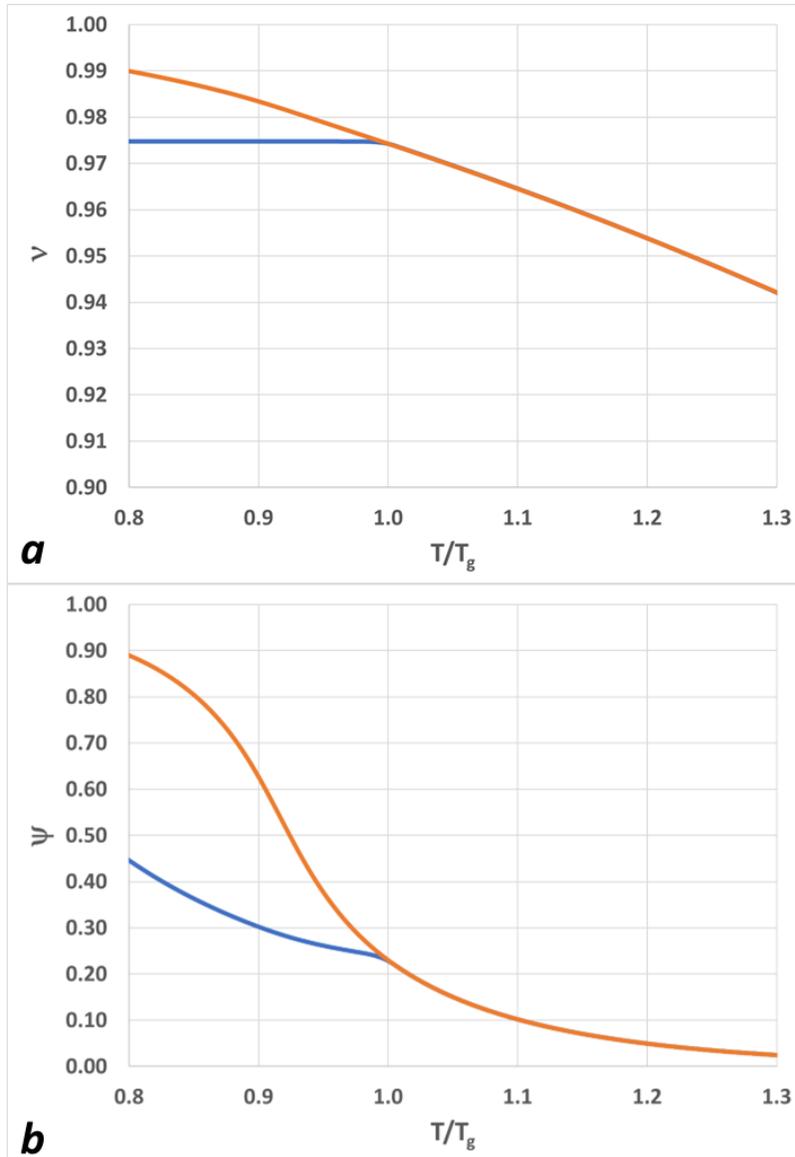

*Figure S 5. The dynamic SL-TS2 variables: (a) $\nu$ (occupancy) and (b) $\psi$ (solid fraction) as a function of normalized temperature, $T/T_g$. The model parameters are summarized in Table 2. The orange curves correspond to equilibrium, and the blue ones correspond to cooling with the rate Q = 1 K/min. The dynamic fragility m = 100. See text for more details.*

Substituting the equilibrium and dynamic solutions shown in Figure S5 into eqs S5b and S6b results in the calculated relaxation times and specific volume shown in Figure 3 of the main text.

## 5. The Relationship between the SL-TS2 and Arrhenius Functions for the α-Relaxation Time

Consider the modified Arrhenius plot for the logarithm of the relaxation time as a function of shifted inverse temperature (Figure S6). The relaxation time is scaled in units of $\tau_{el}$, and the temperature is scaled in units of $T_x$ and shifted by one unit, so that the point X = 0 corresponds to T = $T_x$. The three curves are as follows:

1. Blue line – the numerical SL-TS2 solution described in the previous sections.
2. Orange line – the VFTH solution (eq 11 of the main text).
3. Red line – the high-temperature Arrhenius equation, to be discussed below.

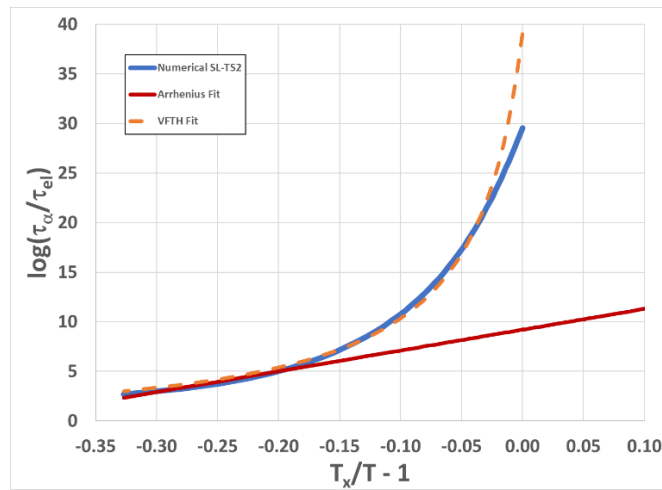

*Figure S 6. The SL-TS2, VFTH, and Arrhenius fit to the relaxation time vs. temperature curve.*

The relationship between the SL-TS2 and VFTH functions has already been discussed in the main text. The Arrhenius curve corresponds to the standard equation, $log\left(\frac{\tau_\alpha}{\tau_{el}}\right) = \frac{E_a}{RT}$, where we assumed $\frac{E_a}{RT_x} = 21.0$, consistent with estimates of Bailly et al.,[6] Schmidtke et al.,[7] and others. Note, however, that in Table S1 (same as Table 2 of the main text), the high-temperature activation energy $\frac{E_1}{RT_x} = 8.0$. The discrepancy between these two numbers comes from the fact that the transition between VFTH and Arrhenius is not sharp, and even the apparently "linear" region still has nonzero fraction of solid elements. Thus, although the slope of the SL-TS2 curve changes in the region -0.35 < (Tx/T − 1) < -0.17, the overall agreement between the Arrhenius function and the SL-TS2 numerical solution is still good.